# A non-invasive investigation of Limoges enamels using both Optical Coherence Tomography (OCT) and spectral imaging: a pilot study

Margaret Read[1,2], Chi Shing Cheung[1], Denise Ling[3], Capucine Korenberg[2], Andrew Meek[2], Sotiria Kogou[1], Haida Liang*[1]

[1] ISAAC, School of Science and Technology, Nottingham Trent University, Nottingham

[2] Department of Scientific Research, The British Museum, London

[3] Department of Conservation, The British Museum, London

*haida.liang@ntu.ac.uk

**ABSTRACT**

This paper investigates the use of Optical Coherence Tomography (OCT) and Short-wave Infrared (SWIR) spectral imaging to study the deterioration of a Limoges enamel panel. Limoges enamels are formed of glass layers applied on a metal substrate and are prone to 'glass disease'. However, the level of deterioration in Limoges enamels is generally difficult to assess visually. In this study, SWIR was used to produce a hydration level map of the enamel, which was coupled with virtual OCT cross-sections. The study shows a good correlation between levels of hydration and structural damage over the enamel panel. Hydration mapping allows visualisation of structural damage across the entire enamel in one image.

## 1   INTRODUCTION

Limoges enamels are a glass-based painted type of enamel from Limoges, France. This type of painted technique applies the enamel layer-by-layer with multiple firings. A typical structure for the painted Limoges enamelling technique starts with a metal substrate, then a first layer of colourless enamel, sometimes a second layer of opaque enamel or a thin foil, then the top layer of either transparent, semi-transparent or opaque coloured enamel.

Limoges enamels from the late 15th and early 16th century seem particularly susceptible to deterioration[1] and many of the enamels show varying degrees of glass disease. On individual pieces there can be a large range in deterioration of the enamel, with difficulty in detecting early stages of the glass deterioration and detailing the exact areas and stage of deterioration unambiguously[2].

Glass deterioration is mainly due to attack by moisture and pollutants in the air. The description of glass deterioration can vary widely within the community, from descriptions such as dulling of the glass to detachment of layers of the glass (spalling). In a review on glass degradation, Kuniki-Goldfinger[2] suggested that the most important information to focus on when conducting a conservation survey on the state of degradation of glass is the formation of salts, the hydration of the glass (gel layers) and crizzling (formation of cracks).

Currently there are many glass studies incorporating a variety of methods, for example, Atomic Force Microscopy[3] (AFM), Scanning Electron Microscope[3] (SEM), Secondary Ion Mass Spectroscopy[3] (SIMS), Raman[4], FTIR[5] and acoustic emissions[5,6]. Although these methods are useful, they are not suitable for large museum surveys as they are either invasive, very time consuming or cannot be done in situ.

## 2  METHODOLOGY

Two analytical methods were employed in situ for this pilot study on a single enamel panel, (a) Optical Coherence Tomography (OCT), which outputs a virtual cross-section image of the subsurface microstructure of the transparent areas of enamel, and (b) Short-wave Infrared (SWIR) spectral imaging used to acquire a spectra of the vibrational modes of water molecules.

### 1.1 Optical Coherence Tomography (OCT)

OCT is based on a Michelson interferometer, which produces a cross-section of the surface and subsurface volume in the form of a 3D image cube. The OCT used in this survey was a ultra-high resolution Fourier Domain OCT at 810nm using a supercontinuum source[7]. This OCT has a depth resolution of ~1.2µm in glass (assuming refractive index n ~1.5), the transverse resolution given by the objective lens is 7µm. The power incident on the object is ~1mW, which has been confirmed to be safe for cultural heritage objects[8]. The image cube collected by the OCT has the dimension 5x5x2mm$^3$ and an acquisition time of approximately 10 seconds.

Ultra-High-Resolution OCT (UHR OCT) is required for the investigation of the deterioration of enamels and glass. The indicators of glass deterioration are the surface deposits and salts that are often <10µm and gel layers close to the surface of the glass. Figure 1 shows the UHR OCT virtual cross-section (B-Scan) image of a deteriorated glass surface (Fig 1a) compared with a 930nm Thorlabs Callisto OCT image of the same glass (Fig 3b). The Callisto OCT has a depth resolution of 4.5µm in glass and transverse resolution of 9µm compared with the depth resolution of 1.2µm for the UHR OCT. The UHR OCT can easily resolve the surface deposits, however, in the lower resolution Callisto OCT, it becomes increasingly difficult to resolve and identify with confidence the deposits that are <15µm.

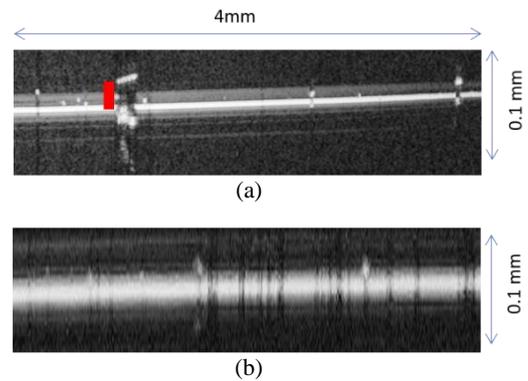

Figure 1 (a) UHR OCT cross-section image of surface deposit, red bar = physical thickness (approx. 24µm) (b) 930nm Callisto OCT cross-section image

OCT measures optical distances, or the time it takes for the light to travel through the medium. Physical depth, or thickness, is measured by dividing the optical thickness by the refractive index. The red bar shown on Fig. 1a is measured from the top of the deposit (air/deposit interface) to the top of the glass (air/glass interface) indicating the physical depth of the surface deposit which is ~24µm. The thickness measured from the top of the deposit to the deposit/bulk glass interface is the optical thickness which is larger than the physical thickness.

OCT is sensitive to the change in refractive index and images the subsurface structure of transparent and semi-transparent materials. Fresnel reflection is strongest at the interfaces with large refractive index changes. Bright, well defined lines denote high reflectivity. For example, the bright interface at the top of the image is the air/glass interface and the bright line towards the bottom of Fig. 2b and 2c are the glass/copper and glass/foil interfaces respectively. Bright, diffused areas denote highly scattering regions, that is regions where there are large differences between the refractive index of the particles and the matrix around them. For example, the white enamel in Fig. 2a. Figure 2 also shows how subsurface make up differs between enamels and how OCT can be used to image these layers.

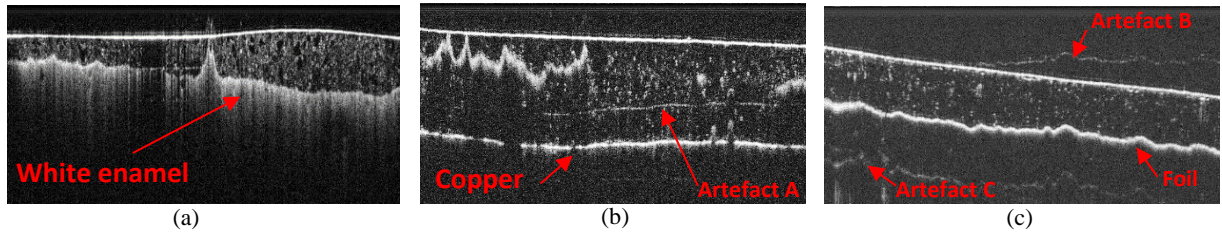

Figure 2. Enamels from Rangers House, English Heritage Collection imaged with the 930nm Callisto OCT (a) Transparent enamel with a white preparation layer (b) Transparent top layer enamel with copper base visible, Artefact A is an auto-correlation artefact (c) Transparent top layer with a foil underneath, Artefact B is an auto-correlation artefact and Artefact C is an inter-reflection artefact

### 1.2 Short-wave Infrared Spectral Imaging

Short-wave Infrared (SWIR) spectral imaging provides spectral and spatial information, which produces an image cube, as seen in Fig. 3. The equipment used for this study was the HySpex SWIR-384 grating-based system, with a spectral range of 1-2.5µm and 288 wavelength bands (spectral sampling interval of 5.45nm). The spatial resolution in this setup is approximately 1mm in both directions. Tungsten Halogen lights were used for illumination of the enamels and the temperature increase was monitored to be <1°C at the object.

SWIR spectral range shows absorption bands corresponding to the vibration modes of molecules. Prominent water absorption bands are located at ~1.4µm, which include the X-OH groups as well as molecular water (H-O-H), and at ~1.9µm which includes just molecular water, H-O-H[9] (Fig. 4). Due to the hydration of glass when it deteriorates, identification of these absorption bands can be linked to how deteriorated the glass is[10]. The spectrum in Fig. 4 is taken from the mulberry area on the 1895.1223.2 enamel (Fig. 5a).

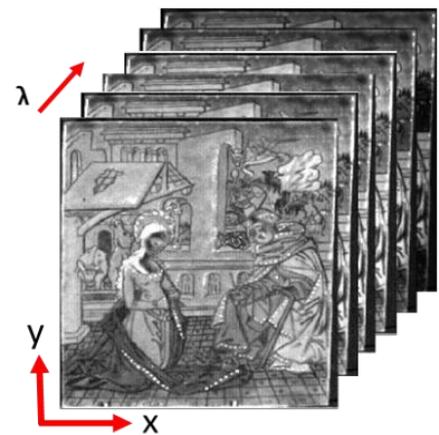

Figure 3. SWIR schematic image cube of enamel 1895.1223.2 stored at the British Museum

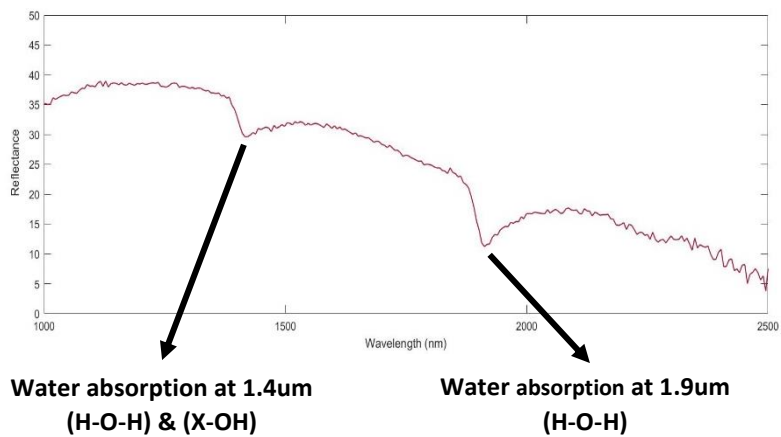

Figure 4. SWIR Spectrum of the mulberry colour of the enamel located on the dress of the lady

# 3 RESULTS

The visual examination of the enamel 1895.1223.2 (Fig. 5a) indicated there was chemical deterioration within the enamel, however, to subdivide the extent and stage of deterioration of affected areas is not possible visually. Figure 5b pictures an angled view of the same enamel to show the diverse range of the deterioration across the glass surface of a single enamel.

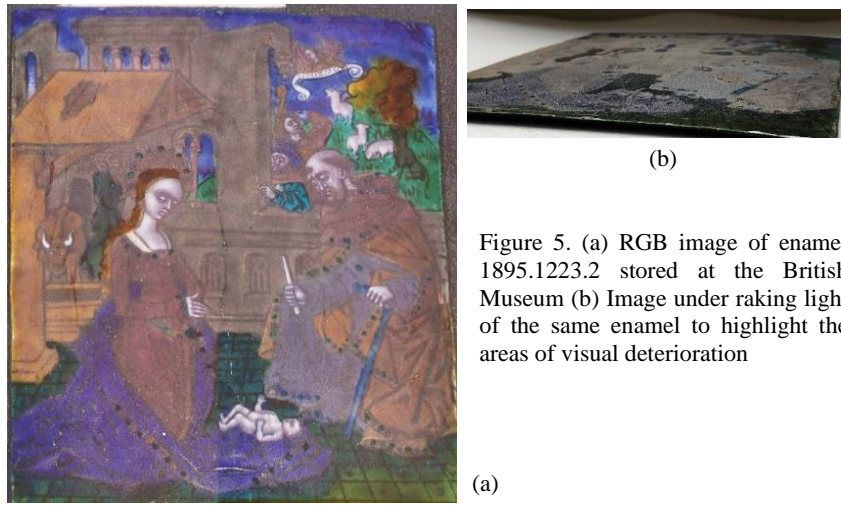

Figure 5. (a) RGB image of enamel 1895.1223.2 stored at the British Museum (b) Image under raking light of the same enamel to highlight the areas of visual deterioration

## 3.1 Hydration Mapping

Hydration mapping is the result of mapping the level of hydration of each pixel over the entire enamel. The level of hydration is calculated by taking the spectrum of each pixel, choosing the absorption band and then taking the Equivalent Width[11] of the chosen band. Figure 6 shows the Hydration Map for enamel 1895.1223.2 where the equivalent width of the 1.9µm water absorption band was used as it gives only the molecular water content. The Hydration Map assessment agrees with what is seen visually but with more information about the differences in hydration of the enamel between the deteriorated areas.

## 3.2 OCT with Hydration Mapping

Hydration mapping coupled with OCT can give both the hydration of the enamel and the structural information of gel layers, surface deposits and crizzling. Figure 7 shows the hydration map of enamel 1895.1223.2 with numbered areas marked to correspond to the OCT image cross-section.

Area 1 on the hydration map (Fig. 7) shows one of the most hydrated areas, this corresponds to the OCT image in Fig. 7.1 which shows evidence of a broken gel layer, surface deposits and crizzling. Area 2 shows an intermediate level of hydration and corresponds to OCT image in Fig. 7.2 which shows evidence of surface deposits. Area 3 shows an area of low hydration and corresponds to the OCT image in Fig. 7.3 showing a mostly smooth air/glass interface with the very low levels of surface deposits. The OCT and the hydration mapping both agree with visual assessments.

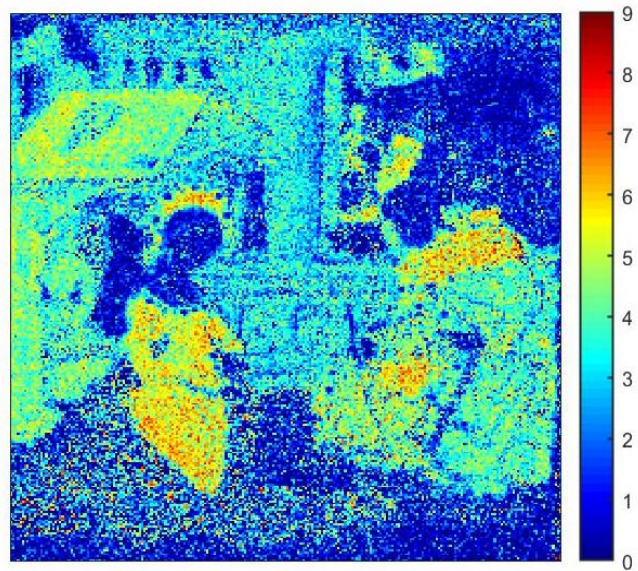

Figure 6. Hydration map at 1.9um of enamel 1220.1223.2, red areas denote a high level of absorption, blue areas denote areas of low absorption

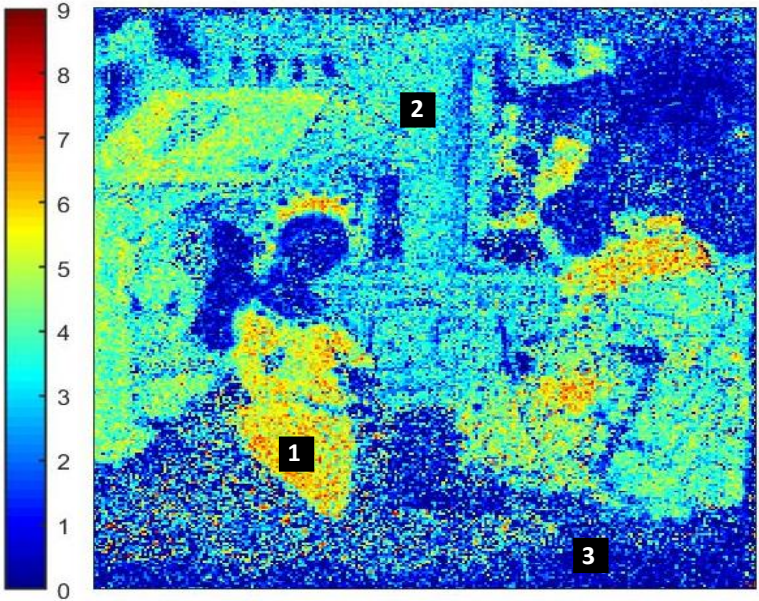

Figure 7. Hydration map at 1.9um of enamel 1220.1223.2, with numbered areas corresponding to the OCT cross-section images below. All OCT images are 5mm (width) x 0.7mm (depth)

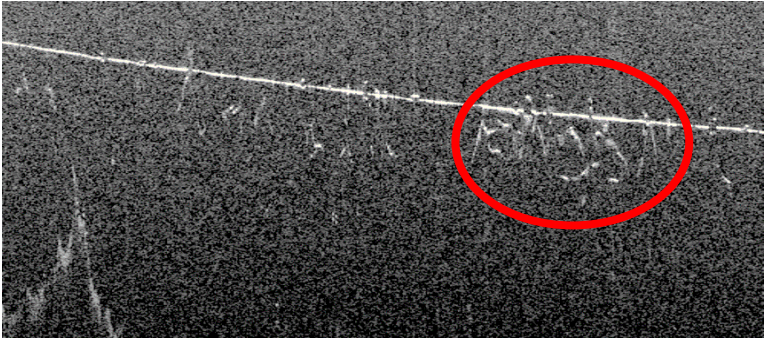

**1.**

The OCT image shows a broken gel layer and surface deposits around the air/glass interface on the OCT cross-section image. The hydration map details this area to be a mix between highly degraded to the most deteriorated on the enamel.

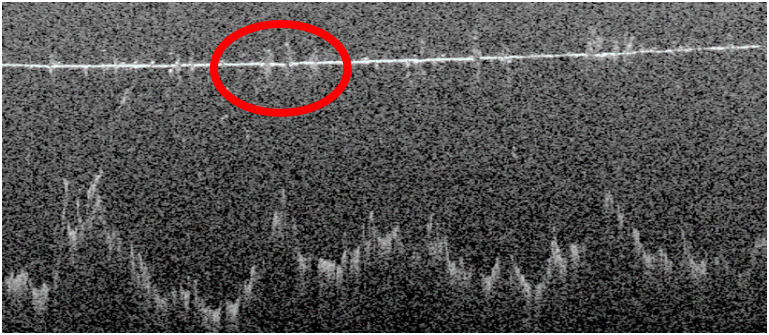

**2.**

The OCT image shows surface deposits at the air/glass interface on the OCT cross-section image. The hydration map details this area to be an area of intermediate degradation.

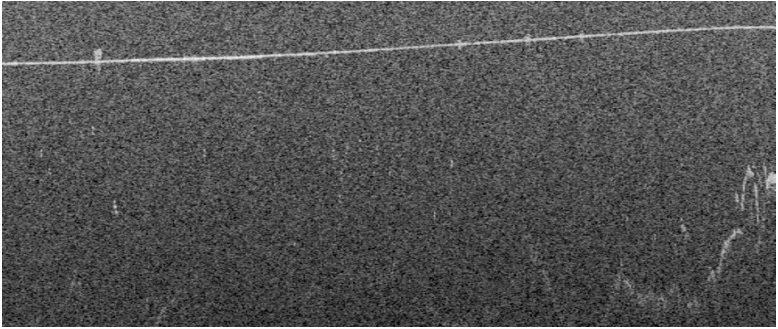

**3.**

The OCT image shows very few surface deposits on the OCT cross-section image. The hydration map details this area to be the least deteriorated area.

## 4 CONCLUSION

OCT coupled with SWIR spectral imaging can be used for preventative conservation to assess the level of deterioration across the entire enamel and identify the enamels most at risk. OCT and SWIR spectral imaging can give detailed, clear-cut information about the hydration and surface and subsurface microstructure of an enamel, both of which are linked closely to glass deterioration. The in situ, rapid scanning of the systems means large museum collections can be assessed on their hydration levels and structural deterioration. Hydration mapping can be used as a standalone, non-invasive, assessment to identify the levels of hydration of an object or as part of multimodal assessment. These techniques can potentially flag enamels that are at risk before the worst visual deterioration has occurred.

## ACKNOWLEDGEMENTS

Funding from the UK Arts and Humanities Research Council (AHRC) Collaborative Doctoral Programme is gratefully received. Thank you to British Museum staff in Scientific Research, Conservation and Britain, Europe and Prehistory departments and those from the NTU ISAAC Lab who supported and helped throughout this pilot study.

## REFERENCES

[1] Smith, R., Carlson, J.H. and Newman, R.M., "An investigation into the deterioration of painted Limoges enamel plaques c. 1470–530," Studies in conservation, 32(3), pp.102-113 (1987)

[2] Kunicki-Goldfinger, J., "Unstable historic glass: symptoms, causes, mechanisms and conservation," Reviews in Conservation, 9, 47-60. 10.1179/sic.2009.54, (2008)

[3] Melcher, M., Wiesinger, R. and Schreiner, M., "Degradation of glass artifacts: application of modern surface analytical techniques," Accounts of chemical research, 43(6), pp.916-926. (2010)

[4] Robinet, L., Coupry, C., Eremin, K. and Hall, C., "Raman investigation of the structural changes during alteration of historic glasses by organic pollutants," Journal of Raman Spectroscopy: An International Journal for Original Work in all Aspects of Raman Spectroscopy, Including Higher Order Processes, and also Brillouin and Rayleigh Scattering, 37(11), pp.1278-1286. (2006)

[5] Thickett, D., Cheung, C.S., Liang, H., Twydle, J., Maev, R.G. and Gavrilov, D., "Using non-invasive non-destructive techniques to monitor cultural heritage objects," Insight-Non-Destructive Testing and Condition Monitoring, 59, 5, pp.230-234. (2017)

[6] Studer, J., "Application of acoustic emission technique to Limoges enamels for damage assessment," In CeROArt. Conservation, exposition, Restauration d'Objets d'Art (No. EGG 2). Association CeROArt asbl. (2012)

[7] Cheung, C.S., Spring, M. and Liang, H., "Ultra-high-resolution Fourier domain optical coherence tomography for old master paintings," Optics express, 23, 8, pp.10145-10157. (2015)

[8] Liang, H., Mari, M., Cheung, C.S., Kogou, S., Johnson, P. and Filippidis, G., "Optical coherence tomography and non-linear microscopy for paintings–a study of the complementary capabilities and laser degradation effects," Optics express, 25, 16, pp.19640-19653. (2017)

[9] Stolper, E., "Water in silicate glasses: an infrared spectroscopic study," Contributions to Mineralogy and Petrology, 81, 1, pp.1-17. (1982)

[10] Buechele, A., Brostoff, L., Zaleski, S., Deems, N., Montagnino, E., Muller, I., Pegg, I., Ward-Bamford, C.L., Loew, M. and Xie, X., "Use of Microscopy and Microanalysis in Assessing Kinetics of Degradation in 19th-century Heritage Glasses" Microscopy and Microanalysis, 24, S1, pp.2138-2139. (2018)

[11] Ridpath, I., "equivalent width." [In A Dictionary of Astronomy], Oxford University Press (2012)